\documentclass[aps,prl,floatfix]{revtex4}
\usepackage{epsfig}

\begin{document}

\title{Line-broadening in gravitational radiation from gamma-ray burst-supernovae}
\author{Maurice H.P.M. van Putten}
%\altaffiliation{Email: marco.cavaglia@port.ac.uk}
\affiliation{LIGO Project, NW17-161, 175 Albany Street, Cambridge, MA 02139}
\author{Hyun Kyu Lee and Chul H. Lee}
\affiliation{Department of Physics, Hanyang University 133-791, Seoul Korea}
\author{Hongsu Kim}
\affiliation{Astronomy Program, SEES, Seoul National University,
Seoul 151-742, Korea}

\begin{abstract}
Core-collapse in massive stars is believed to produce a rapidly
spinning black hole surrounded by a compact disk or torus. This
forms an energetic MeV-nucleus inside a remnant He-core, powered
by black hole-spin energy. The output produces a GRB-supernova,
while most of the energy released is in gravitational radiation.
The intrinsic gravitational-wave spectrum is determined by
multipole mass-moments in the torus. Quadrupole gravitational
radiation is radiated at about twice the Keplerian frequency of
the torus, which is non-axisymmetric when sufficiently slender,
representing a ``{\em black hole-blob}" binary system or a ``{\em
blob-blob}" binary bound to the central black hole.  We here
discuss line-broadening in the observed spectrum due to
Lense-Thirring precession, which modulates the orientation of the
torus to the line-of-sight. This spectral feature is long-lived,
due to weak damping of precessional motion. These events are
believed to occur perhaps once per year within a distance of
100Mpc, which provides a candidate source for Advanced LIGO.
\end{abstract}

\preprint{LIGO-P030026-00-R}


\maketitle

LIGO recently completed its second science run \cite{laz03}. The
results demonstrate the feasibility of km-sized laser interferometric
gravitational-wave detectors. In the broad band width of 20-1000Hz,
advanced LIGO is sensitive to extragalactic events of binary neutron-star
coalescence and GRB-supernovae from rotating black holes. This band width
also covers emissions from new born neutron stars, binary coalescence of
black holes, as well as emissions from rapidly rotating neutron stars.
Anticipating the spectrum of gravitational radiation from candidate
sources is important in designing optimal detection strategies. While
the spectrum of binary inspiral is well-understood to high post-Newtonian
order at large binary separations, that of GRB-supernovae is only
beginning to be identified.

Core-collapse of massive stars is considered a potential site of
gravitational radiation \cite{nak89,moe91,bon94,dav02,fry02,min02,kob02}.
Here, gravitational radiation is produced by the release of gravitational
binding energy during collapse and in accretion processes on a newly
formed black hole (e.g, \cite{fry01}). If accretion is driven by
angular momentum loss in magnetic winds, the timescales in core-collapse
by magnetic regulated hyperaccretion are relatively short: about 0.05s
for the outer accretion disk, and about 0.01s for the inner accretion
disk \cite{mvp01a}. Hyperaccretion flows are probably strongly turbulent,
which would imply a broad gravitational-wave spectrum.
Aforementioned studies appear to indicate an energy output, which leaves
a range of detectability by ground based detectors of up to about 10Mpc.
These events should therefore be considered in the context of core-collapse
events independent of the GRB phenomenon, in light of current estimates on
the local GRB event-rate of about one per year within a distance of 100Mpc
\cite{fra01,mvp03c}. These studies on gravitational radiation in
core-collapse of a massive star do not invoke the spin-energy of
a newly formed black hole.

The lifetime of rapid spin of the black hole surrounded by a torus in
suspended accretion is about tens of seconds \cite{mvp01a,mvp04}. This
represents the timescale of ``unseen" dissipation of black hole-spin energy
in the event horizon, the rate of which is
governed by the poloidal magnetic field-energy in the surrounding torus
\cite{mvp01a,mvp04}. This timescale is in good agreement with the durations of
long GRBs, which reflect the lifetime of activity of the
inner engine \cite{pir98}.

Optimal detection strategies for gravitational radiation from GRBs may
be facilitated by a priori knowledge from a specific model. Our model
for GRB-supernovae from rotating black holes describes the formation
of a GRB accompanied by a radiatively-driven supernova, while most of the
energy output is in gravitational radiation \cite{mvp04}. The lowest order
quadrupole moment is manifest as a
``{\em black hole-blob}" binary system or a ``{\em blob-blob}" binary bound to
the central black hole when
the torus is sufficiently slender, giving rise to \cite{mvp04}
\begin{eqnarray}
E_{gw}\simeq 4\times 10^{53}\mbox{erg}
\left(\frac{\eta}{0.1}\right)\left(\frac{M_H}{7M_\odot}\right),~~~
f_{gw}\simeq 500\mbox{Hz}
\left(\frac{\eta}{0.1}\right)\left(\frac{M_H}{7M_\odot}\right)^{-1}
\label{EQN_EGW}
\end{eqnarray}
in gravitational radiation. Here, $\eta=\Omega_T/\Omega_H$ denotes
the ratio of the angular velocity $\Omega_T$ of the torus
to the angular velocity $\Omega_H$ of a rapidly rotating
black hole of mass $M_H$. This energy output (\ref{EQN_EGW})
results from a causal spin-connection to the black hole
of a surrounding torus with multipole mass-moments.
For an intrinsic mass-inhomogeneity $\delta M_T$ in the torus, we have
a luminosity of gravitational radiation according to
\begin{eqnarray}
L_{gw}=({32}/{5})(M_H/R)^5(\delta M_T/M_H)^2,
\end{eqnarray}
where ${\cal M}\simeq M_H(\delta M_T/M_H)^{3/5}$ denotes the chirp
mass. The relative mass-inhomogeneity $\delta M_T/M_H$ is
determined self-consistently by the balance equations of energy
and angular momentum flux in a suspended accretion state to be
about a few promille \citep{mvp04}. The fractions of black
hole-spin energy radiated in various channels is independent of
the mass of the torus (provided $M_T$ is a few tenths of a solar
mass). The mass of the torus only affects the lifetime of rapid
spin of the black hole and hence the duration of the burst.

The energy in gravitational waves (\ref{EQN_EGW}) is larger than the true
energy $E_\gamma$ in GRB-afterglow emissions \cite{mvp04} (see further \cite{bet03}),
i.e.:
\begin{eqnarray}
E_{gw}\simeq 10^3 E_\gamma.
\end{eqnarray}
We attribute the GRB-afterglow emissions to baryon poor outflows
along an open magnetic flux-tube subtended at a finite opening
angle on the event horizon of the black hole. Statistical analysis
of GRBs with individually measured redshifts points towards
strongly beamed GRB-emissions \cite{fra01} in the form of a highly
anisotropic beam accompanied by very weak GRB-emissions over
essentially all angles \cite{mvp03c}. This is distinct from purely
conical outflows with uniform emissions over a finite opening
angle and zero emissions outside. The non-uniform strongly
anisotropic beam gives rise to an anti-correlation between
inferred opening angle and redshift, given a standard true
GRB-energy output in response to a finite flux-limit of a
detector. The weak ``all-angle" emissions are proposed to be at
the level of GRB980425, or about $10^{-4}$ times the standard GRB
luminosities -- weak, but significant in allowing nearby GRBs to
be identified off-axis as apparent anomalous, low-luminosity
events.

A radiatively-driven supernova energy accompanies the GRB and burst in
gravitational waves powered by magnetic winds
coming off the torus in suspended accretion around the central black hole
\cite{mvp04}. The predicted kinetic energy
\begin{eqnarray}
E_{SN}\simeq 2\times 10^{51}\mbox{erg}
\end{eqnarray}
is in good agreement with the
observed energy of about 2$\times10^{51}$erg in the aspherical explosion
of SN1998bw \cite{hoe99}. We note that the ``hypernova" energies
\cite{pac98} of a
few times $10^{52}$ erg \cite{iwa98,woo99} are isotropic equivalent
energies assuming spherical symmetry, not true kinetic energies in
the aspherical explosions.

We here study the spectral signature of Lense-Thirring precession
in (\ref{EQN_EGW}). Lense-Thirring precession can be described by
nodal precession \cite{wil72}. In the slow-rotation limit, its
angular velocity $\Omega_{LT}$ \cite{len18} agrees essentially to
the frame-dragging angular velocity. The orientation of the torus
is well-defined, in that misalignment between its angular momentum
and angular velocity is of second order, $O(\theta\Omega_{LT})$,
where $\theta$ denotes the wobbling angle. This contrasts with
rigid body motion, where precession is due to a misalignment
between the angular momentum and angular velocity vectors.
Lense-Thirring precession is frequently invoked in modelling QPOs
produced by accretion disks in X-ray binaries \cite{ste99}. It
represents a general relativistic effect due to frame-dragging,
and hence it acts universally on the large-scale structure of a
torus surrounding a black hole. A torus which is misaligned with
the spin-axis of the black hole precesses with essentially the
frame-dragging angular velocity described by the Kerr metric. In
Boyer-Lindquist coordinates, we have $\Omega_{LT}\simeq
{2J_H}/{R^3}$ for a black hole angular momentum
$J_H=M^2\sin\lambda$ in terms of the mass $M_H$ and the specific
angular momentum $\sin\lambda=a/M$. Given the angular velocity
$\Omega_H=\tan(\lambda/2)/2M$ of the black hole and the angular
velocity $\Omega_T\simeq M^{1/2}R^{-3/2}$ of the torus, we have
\begin{eqnarray}
\frac{\Omega_{LT}}{\Omega_H}\simeq 2\times 10^{-2}\left(\frac{\eta}{0.1}\right)^2
\sin^2(\lambda/2).
\label{EQN_OMH}
\end{eqnarray}
At nominal values $\eta\sim 0.1$, $\Omega_{LT}$ is about 10\% of $\Omega_T$, or, equivalently,
about 1\% of $\Omega_H$. The associated precession of the black hole, due to
conservation of total angular momentum, can generally be neglected
when the angular momentum of the torus is much less than that of
the black hole.

Our study considers Lense-Thirring precession in response to an
initial misalignment of the torus relative to the spin-axis of the
black hole. The active nucleus is believed to form in
core-collapse of a massive stars \cite{woo93} {\em inside} a
remnant envelope (e.g. \cite{mvp04}). The angular momentum of the
black hole is commonly attributed to orbital angular momentum,
received during spin-up in a common envelope phase with a
companion star \cite{pac98,bro00}. A misalignment in the nucleus
may hereby reflect misalignment between the spin of the progenitor
star and the orbital motion of the companion star. Lense-Thirring
precession modulates the orientation of the torus to the
line-of-sight, introducing phase-modulation in the line-emissions
as observed at the detector. This broadens the emission lines, and
may produce side-bands similar to those arising in amplitude
modulation. There arises an additional pair of low similar to
radiation coming off a freely precessing neutron star. These
low-frequency emissions produce relaxation, and gradually bring
the torus into alignment. Their low-luminosities allow the
misalignment and its precession to be long-lived.

For the two polarizations of gravitational waves, we have
\begin{eqnarray}
h_+=2(1+\cos^2\iota)\cos(2\Omega_T t),~
h_\times=-4\cos\iota\sin(2\Omega_T t)
\label{EQN_H3}
\end{eqnarray}
where $\iota$ denotes the angle between the angular momentum and the
the line-of-sight. Here, we have suppressed a coefficient
$2\left({e_k}/{M_H}\right)\left({M_H}/{r}\right),$
where $e_k$ denotes the kinetic energy of the mass-inhomogeneity $m$ in the torus,
and $r$ denotes to the distance to the source.
Precession of the torus introduces
a time-varying angle $\iota(t)$, which modulates the strain amplitudes
$h_+$ and $h_-$ at the observer (Fig. 1).
Given a mean angle $\iota_0$ of the angular momentum of the torus
to the line-of-sight and a wobbling angle $\theta$,
the time-dependent angle $\iota(t)$ of the same satisfies
\begin{eqnarray}
\cos\iota(t)=\sin\iota_0\sin\theta \cos(\Omega_{LT}t) + \cos\iota_0\cos\theta.
\end{eqnarray}
Substitution of $\cos\iota(t)$ into (\ref{EQN_H3}) produces
phase-modulation at the Lense-Thirring frequency.
The expressions (\ref{EQN_H3}) hereby produce
line-broadening by phase-modulation about the carrier frequency
$2\Omega_T$, according to
\begin{eqnarray}
h_+=6\cos(2\Omega_Tt)+\cos(2\Omega_Tt+2\iota)+
\cos(2\Omega_Tt-2\iota),
\end{eqnarray}
and
\begin{eqnarray}
h_\times= -2\left[\sin(2\Omega_Tt+\iota)+\sin(2\Omega_Tt-\iota)\right].
\end{eqnarray}
The resulting wave-forms are illustrated in Fig. 2.

In case of $\theta<<\iota_0$, we may linearize (\ref{EQN_H3})
$h_+  =h_+^{(0)}+\theta h_+^{(1)} +O(\theta^2),$
$h_\times  =h_\times^{(0)} +\theta h_\times^{(1)} + O(\theta^2),$
where $h_+^{(0)}=h_+(\iota_0)$ and $h_\times^{(0)}=h_\times(\iota_0)$
refer to the strain-amplitudes (\ref{EQN_H3}) with $\iota=\iota_0$,
and $h_+^{(1)}
=-\sin(2\iota_0)
\left[\cos([2\Omega_T+\Omega_{LT}]t)+\cos([2\Omega_T-\Omega_{LT}]t)\right],$
and $h_\times^{(1)}
=2\sin(\iota_0)
\left[\sin([2\Omega_T+\Omega_{LT}]t)+\sin([2\Omega_T-\Omega_{LT}]t)\right].$
For small wobbling angles, therefore,
phase-modulation reduces to amplitude
modulation, producing a carrier frequency accompanied by two side-bands
at frequencies
\begin{eqnarray}
2\Omega_T-\Omega_{LT},~~~2\Omega_T,~~~2\Omega_T+\Omega_{LT}.
\end{eqnarray}
The ratio of the amplitude line-strengths of the side bands to that of the
carrier satisfies
\begin{eqnarray}
K\simeq \theta\left(\frac{1+\cos^2\iota_0}
{1+6\cos^2\iota_0+\cos^4\iota_0}\right)^{1/2}
\sin\iota_0,
\end{eqnarray}
where we used $\Omega_{LT}<<2\Omega_{T}$. Averaged over all angles
$\iota_0$, we have $\bar{K}\simeq \theta/2$.
A wobbling angle of about 30$^o$ typically produces
side-bands of relative strength $20\%$ (taking together
$h_+$ and $h_\times$ in each side-band).

Low-frequency emission lines are found by considering the projections
of a precessing torus onto the
celestial sphere, as shown in Fig. 1. Translating the geometrical picture of
\citep{zs79} to the case of a torus around a black hole, we
observe that (1) projections along the spin-axis of the black hole
on the celestial sphere produces an ellipse which rotates at twice
the Lense-Thirring frequency; (2) projections of the torus along
directions in the orbital plane on the celestial sphere produce a
line-segment which oscillates at the Lense-Thirring frequency. A
precessing torus hereby introduces a spectral anisotropy,
whereby emissions at twice the precession frequency are
preferentially along the spin-axis of the black hole, and
emissions at the precession frequency are preferentially along
directions in the equatorial plane.

The strain-amplitude of the low-frequency emissions satisfies \citep{zs79}
\begin{eqnarray}
{\bf h}_{LT} =
\sin(2\theta)\frac{\Omega_{LT}^2\Delta I}{2r}
\hat{{\bf h}}_{LT},
\end{eqnarray}
where
\begin{eqnarray}
\hat{{\bf h}}_{LT}=
{\bf e}_+ \left(h^{||}_+\tan\theta + h^{\perp}_+\right) +
{\bf e}_\times \left(h^{||}_{\times}\tan\theta +
h^{\perp}_{\times}\right)
\label{EQN_hLT}
\end{eqnarray}
and
\begin{eqnarray}
h^{||}_+     ={2(1 + \cos^2\iota)}\cos(2\Omega_{LT}t),
h^{||}_{\times}   ={4\cos\iota}\sin(2\Omega_{LT}t),\\ h^{\perp}_+
=\frac{1}{2}\sin(2\iota)\cos(\Omega_{LT}t),
h^{\perp}_{\times}=\sin\iota\sin(\Omega_{LT}t).
\label{EQN_SP}
\end{eqnarray}
The amplitude ratio of (\ref{EQN_hLT}) relative to
(\ref{EQN_H3}) satisfies
\begin{eqnarray}
\frac{h_{LT}}{h}\simeq \frac{1}{8}\left(\frac{\Omega_{LT}}{\Omega_T}\right)^2
\left(\frac{m}{M_T}\right)^{-1}<1\%.
\end{eqnarray}
Here, we note that $M_T/m$ is a ratio of order unity, for $m$ on
the order of a few permille of $M_H$ in a torus of a few tenths of
$M_\odot$. Emissions by (\ref{EQN_hLT}) along the axis are second
order in the wobbling angle $\theta$; emissions in directions
along the orbital plane are first order in $\theta$. This reflects
the amplitude of the time-variability of the quadrupole moment in
the association projections on the celestial sphere. The
associated net luminosity satisfies
\begin{eqnarray}
L^{LT}_{gw}= \frac{1}{10}\Omega_{LT}^6 (\Delta I)^2
                \sin^2(2\theta)(1 + 16 \tan^2 \theta),
\label{EQN_LGW}
\end{eqnarray}
where the first and second part in the bracket correspond to the
emissions by $h^{\perp}$ and $h^{||}$, respectively at frequencies
$\Omega_{LT}$ and $2\Omega_{LT}$. Their contributions are equal
when $\tan\theta=1/4$ or $\theta\simeq14^o$, and the emissions in
$h^{\perp}$ dominate for wobbling angles $\theta<14^o$. In this
event, the spectrum produced by (\ref{EQN_hLT}) varies according
to the different dependencies on the viewing angle $\iota$.
Viewing the source edge on (the observer in the orbital plane)
sees $h^{||}_+$ at $2\Omega_{LT}$ and $h^{\perp}_{\times}$ at
$\Omega_{LT}$. Observers along the spin-axis of the central black
hole see no lines at $\Omega_{LT}$ in $h^{\perp}_+$ and
$h^{\perp}_{\times}$, but a single line at $2\Omega_{LT}$ in both
polarizations of $h^{||}$ which are second-order in $\theta$.

The lifetime of precession can be calculated by considering the
backreaction of gravitational radiation on the torus.
Gravitational radiation backreaction forces consist of dynamical
self-interactions and radiation-reaction forces \citep{tho69,
cha70,shu80}. For sources with nonrelativistic motions in the
approximation of weak internal gravity, the latter can be modelled
by the Burke-Thorne potential as it arises in the $2\frac{1}{2}$
post-Newtonian approximation. There is no change  in the
continuity equation in this intermediate order (such as in the
second-order post-Newtonian approximation
\citep{cha70,shu80,shu83}).

Gravitational radiation backreaction on an intrinsic quadrupole
moment produced by the $m=2$ Papaloizou-Pringle instability, is
governed by the small parameter
$\beta = ({\pi R\Sigma}/{10})\left(2\Omega_TR\right)^5$, or
\begin{eqnarray}
\beta =
5\times10^{-5}\left(\frac{\eta}{0.1}\right)^{5/3}
\left(\frac{M_T}{0.1M_\odot}\right)
\left(\frac{M_H}{b}\right)
\left(\frac{7M_\odot}{M_H}\right),
\end{eqnarray}
where $b\sim M$ denotes the minor radius of the torus in suspended
accretion. This backreaction {\em stimulates} the $m=2$ instability
\citep{mvp02a}.

Gravitational radiation backreaction to
the precessional motion is due to the low-frequency emissions, since
the emissions due to intrinsic multipole mass-moments
leave the wobbling angle unaffected.
The backreaction of the low-frequency emissions are similar but not
identical to that on a freely precessing neutron star.
Adapting \cite{cut00} to the present case,
the low-frequency gravitational radiation reaction
torque $T_{gw}^{LT}$
is perpendicular to the symmetry axis of the torus, along
which there exists a discrete symmetry in the presence of
intrinsic multipole mass-moments.
Gravitational radiation reaction torques associated with
the latter act along this symmetry axis.
It extracts the angular momentum along the spin-axis of the
black hole -- the axis about which the torus wobbles -- and
angular momentum associated with the wobble angle itself. In the suspended
accretion state, the angular momentum along the spin-axis of
the black hole is constant on timescales less than the lifetime
of rapid spin of the black hole -- the GRB timescale of tens of
seconds. We may follow \cite{cut00}, by imposing the constraint
\begin{eqnarray}
J_z=\mbox{const.}
\label{EQN_C}
\end{eqnarray}
Reduction of the wobbling angle reflects
extraction of the kinetic energy of the wobbling
motion of the torus. According to \citep{cut00}, we have
\begin{eqnarray}
\dot{\theta}=-\frac{T_{gw}^{LT}\cos\theta}{J},
\end{eqnarray}
where $T_{gw}^{LT}\Omega_{LT}\sin\theta=L_{gw}^{LT}$ \cite{cut00}.
At late times, wobble angle is small, whereby
$\dot{\theta} = - {\theta}/{\tau_{\theta}}$
with $\tau_{\theta}^{-1} =
({1}/{10})\left({\Omega_{LT}}/{\Omega_H}\right)^4
\Omega_H^4M_TR^2$ for a torus of mass $M_T$ and major radius $R$.
In dimensional units, and using (\ref{EQN_OMH}), we have
\begin{eqnarray}
\tau_\theta\simeq 2\times
10^{6}\left(\frac{\eta}{0.1}\right)^{-28/3}
\left(\frac{M_H}{7M_\odot}\right)\mbox{s}.
\end{eqnarray}
For a wide range of values of $\eta$ and $M_H$, the
damping time due to the radiation backreaction is
longer than the time-scale of tens of seconds of long GRBs.

We may compare the strength of the high and low-frequency emissions.
The luminosity at frequency $2\Omega_T$ due to an intrinsic quadrupole
moment in the torus of mass $M_T$ satisfies
\begin{eqnarray}
L_{gw}(2\Omega_T)=3\times 10^{-4}\left(\frac{\eta}{0.1}\right)^{10/3}
                   \left(\frac{m}{M_T}\right)^2.
\end{eqnarray}
The luminosity in emissions of $h^{\perp}$ associated with a small wobbling angle $\theta$
satisfies
\begin{eqnarray}
L_{gw}(\Omega_{LT})=1.6\times 10^{-9}\left(\frac{\eta}{0.1}\right)^{10/3}
                       \left(\frac{M_T}{M_H}\right)^2\theta^2,
\end{eqnarray}
where we used $\Omega_{LT}/\Omega_T\simeq \eta$. In the suspended accretion
state, we expect a lumpiness $m\simeq 0.15\%M_H$ or larger, whereby
\begin{eqnarray}
\frac{m}{M_T}\simeq 10\%\left(\frac{M_H}{7M_\odot}\right)\left(\frac{0.1M_\odot}{M_T}\right).
\end{eqnarray}
It follows that
\begin{eqnarray}
L_{gw}(\Omega_{LT}):L_{gw}(2\Omega_{T}) = 5\times 10^{-6}
\left(\frac{\theta}{0.1}\right)^2
\left(\frac{m}{0.1M_T}\right)^{-2}.
\end{eqnarray}
We conclude that the low-frequency emissions are low-luminosity
lines, relative to the expected emissions from an intrinsic quadrupole
moment. This
low-luminosity is due to the strong frequency dependence of gravitational
radiation, and the limited source of (kinetic) energy in the wobbling motion
of the torus. In contrast, the gravitational radiation produced by the
intrinsic quadrupole moment is produced by the Keplerian frequency and
it derives from the large reservoir in spin-energy of the the black hole.

%\section{Summary and Discussion}

To summarize, we predict
long bursts of gravitational radiation in GRB-supernovae from rotating
black holes with an energy output of $4\times 10^{53}$erg. These bursts
represent the catalytic conversion of black hole-spin energy into
gamma-rays and gravitational waves, and an accompanying supernova.
The gravitational wave-luminosity is dominated by emissions from multipole
mass-moments in the torus for sufficiently slender tori
(minor-to-major radius of the torus less than about 0.75 for a mass-moment
$m=1$, and less than about 0.33 for mass-moment $m=2$ \cite{mvp04}).
The duration is set by
the lifetime of rapid spin of the black hole. A misalignment of the torus and the
black hole conceivably represents a remnant feature of the angular momentum
distribution in the progenitor binary of a massive star and its companion. It
produces phase-modulation in the observed strain-amplitude.

A detailed understanding of the wave-form of gravitational radiation is important
in applying matched filtering. For matched filtering, we have the signal-to-noise
estimate \cite{mvp04}
\begin{eqnarray}
\left(\frac{S}{N}\right)_{mf}
\simeq 8 \left(\frac{S_h^{1/2}(500\mbox{Hz})}{4\times 10^{-24}\mbox{Hz}^{-1/2}}\right)^{-1}
\eta_{0.1}^{-3/2}M_7^{5/2}d_8^{-1}.
%\left(\frac{\eta}{0.1}\right)^{-3/2}\left(\frac{M_H}{7M_\odot}\right)^{5/2}
%\left(\frac{d_L}{140\mbox{Mpc}}\right)^{-1}.
\label{EQN_SN1}
\end{eqnarray}
Presently, the second science run in LIGO shows
${S_h^{1/2}(500\mbox{Hz})}=5\times 10^{-22}$Hz$^{-1/2}$
\citep{laz03}, which corresponds to $S/N= 8$ for a source at
$d_L\simeq1$Mpc. This will improve to a sensitivity range of
100Mpc with the Advanced LIGO noise-strain amplitude target
$S_h^{1/2}(500\mbox{Hz})=4\times 10^{-24}\mbox{Hz}^{-1/2}$. This
sensitivity range corresponds a true event-rate (seen in
gravitational radiation; seen and unseen in GRB-emissions) of one
per year, based on a GRB beaming factor of about 500
\citep{fra01,mvp03c}.

\acknowledgements
The authors thank the LSC and the anonymous referee for a constructive review.
This work is supported by grant No. 1999-2-11200-003-5 from the
Korea Science \& Engineering Foundation, the BK21 Project
of the Korean Government, and by the LIGO Observatories, constructed
by Caltech and MIT with funding from NSF under cooperative agreement
PHY 9210038.
The LIGO Laboratory operates under cooperative agreement
PHY-0107417. This paper has been assigned LIGO document number
LIGO-P030026-00-R.

\newpage
\begin{figure}
\epsfig{file=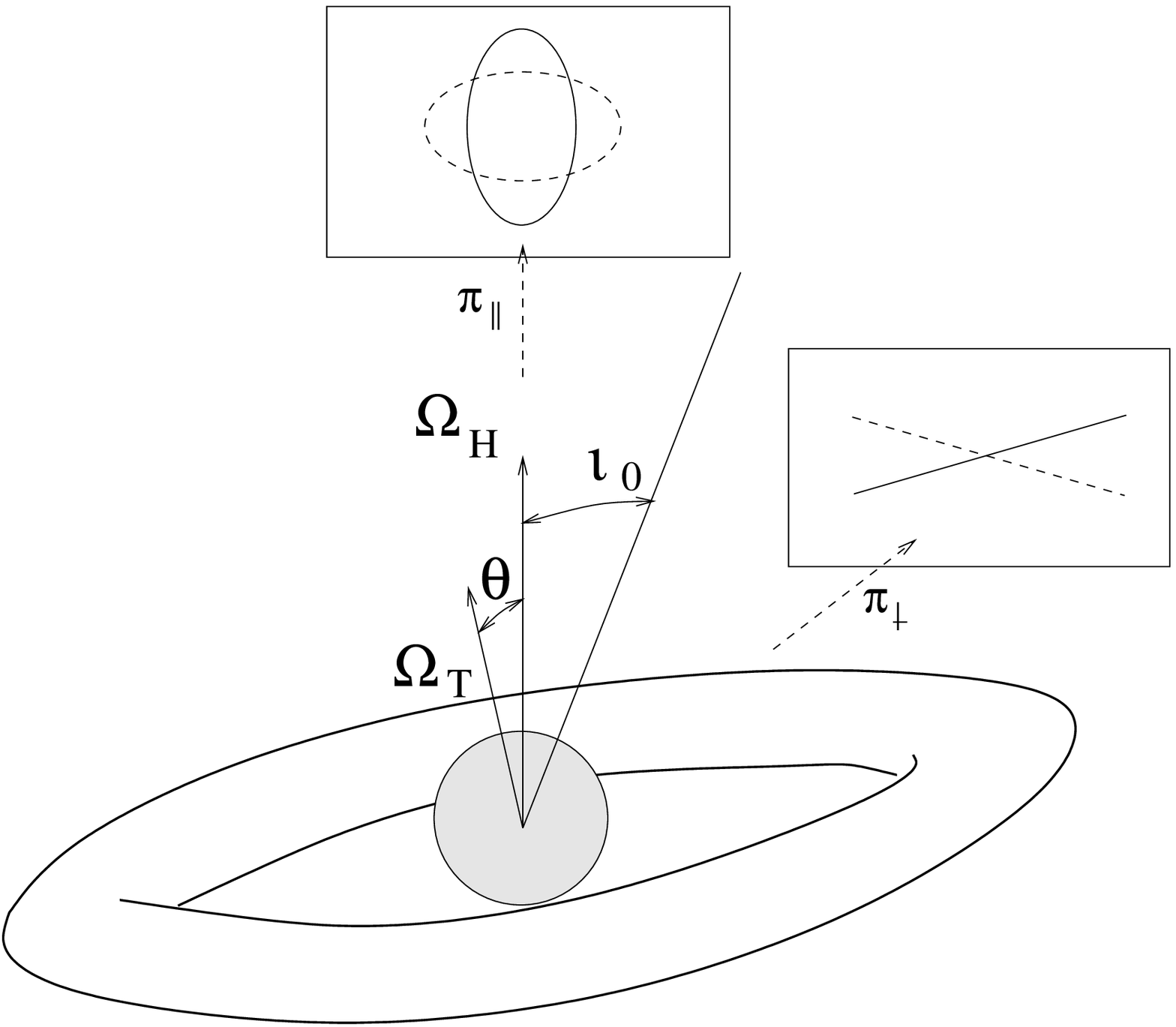,width=100mm,height=80mm}
\caption{The geometry of Lense-Thirring precessing torus with angular velocity
$\Omega_T$ around a rotating black hole with angular velocity $\Omega_H$.
The wobbling angle is $\theta$. The line-of-sight has an angle $\iota_0$
relative to the spin-axis of the black hole. The angle of the orientation
of the torus to the line-of-sight varies in time, between
$\iota_0-\theta$ and $\iota_0+\theta$. This introduces phase-modulation
in the observed strain-amplitude of emissions by an intrinsic
quadrupole moment in the torus. Low-frequency spectral anisotropy is produced by
apparent quadrupole moments, corresponding to projections $\pi_{||}$ and $\pi_{\perp}$
of the torus onto the celestial sphere in directions along and orthogonal to the
spin-axis of the black hole.
The associated modulations are at twice and, respecively, once the Lense-Thirring
precession frequency.}
\end{figure}

\begin{figure}
\epsfig{file=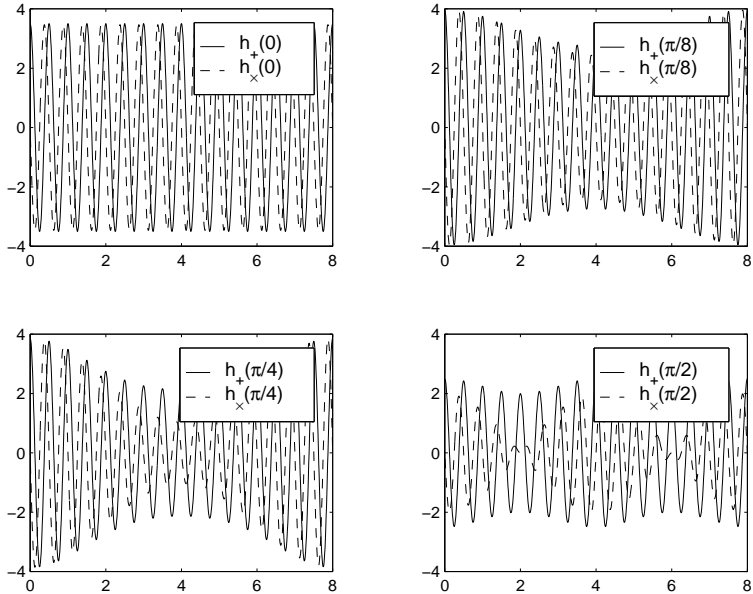,width=100mm,height=80mm}
\caption{The strain-amplitudes $h_+$ and $h_\times$ around twice the
Keplerian angular frequency $\Omega_T$ are produced by
a quadrupole mass-moment in the torus in a state of suspended accretion. The
observed strain-amplitudes are subject to phase-modulation by Lense-Thirring
precession of the torus. For wobbling angles which are small relative to the
mean angle to the line-of-sight, the resulting amplitude modulation produces
two side-bands to the carrier frequency $2\Omega_T$, separated by the
Lense-Thirring frequency $\Omega_{LT}$. Shown are the strain-amplitudes
for the case of $\iota_0=0,\pi/8,\pi/4,\pi/2$ for a wobbling angle of
30$^o$ and a Lense-Thirring precession frequency of 1/8 of the Keplerian
frequency. The amplitude corresponds to a source at unit distance,
and the index refers to the number of Keplerian periods. Additionally, small
contributions to the strain amplitude arise at once and twice $\Omega_{LT}$
in response to apparent quadrupole moments in the projections of
the torus onto the celestial sphere.}
\end{figure}

\end{document}